\documentclass[12pt]{iopart}

\usepackage{graphicx}
\usepackage{hyperref}
\usepackage{multirow}
\usepackage{tabularx}
\usepackage{array}
\usepackage{enumitem}

\usepackage{color}
\usepackage{rotating}
\usepackage[letterpaper,margin=1in]{geometry}
\usepackage{tablefootnote}

\begin{document}

\title[Liu et al.]{Integrating noise into PhET simulations to promote student learning of measurement uncertainty}

\author{Qiaoyi Liu$^{1,2}$, Matthew Blackman$^{2,3}$, Gayle Geschwind$^{1,2}$, Catherine Carter$^{2,3}$, Katherine K. Perkins$^{2,3}$ and H. J. Lewandowski$^{1,2}$}
\address{$^{1}$JILA, National Institute of Standards and Technology and University of Colorado Boulder, Boulder, Colorado 80309, USA}
\address{$^{2}$Department of Physics, University of Colorado Boulder, Boulder, Colorado 80309, USA}
\address{$^{3}$PhET Interactive Simulations, University of Colorado Boulder, Boulder, Colorado 80309, USA}
\ead{qili7958@colorado.edu}
\vspace{10pt}

\begin{abstract}
Understanding concepts and practices of measurement uncertainty is a core competency of physicists and engineers, and many physics lab courses aim to have students learn these ideas. However, there is strong evidence that these goals are often not met. To address the challenge of improving students’ proficiency with measurement uncertainty concepts and practices, we designed and developed a new PhET simulation, \textit{Projectile Data Lab} (PDL), featuring statistical noise and measurement tools in the context of projectile motion. We integrated this simulation into the Common Online Data Analysis Platform (CODAP), creating an instructional platform for collecting and analyzing data from the simulation, and designed three simulation-based instructional activities for instructors to use in their lab courses. We describe the pedagogical design of the new simulation, the PDL+CODAP instructional platform, and the associated instructional activities. We highlight how the targeted learning goals guided the pedagogical design, as well as how these three instructional tools (the simulation, the PDL+CODAP platform, and the lab activity) work together and leverage the affordances of each to scaffold learning. The goal of this work is to provide a model of how noise-enhanced simulations and activities can be designed to enhance student learning of measurement uncertainty. 
\end{abstract}

%
\vspace{2pc}
\maketitle
%
%

\section{Introduction}
The PhET Interactive Simulations team at University of Colorado Boulder has been researching and improving the design and classroom use of interactive simulations to enhance physics teaching and learning for more than 20 years~\cite{perkins2006phet, finkelstein2006high, wieman2008phet, wieman2010teaching, banda2021effect}. Using implicit scaffolding, each PhET simulation creates an intuitive exploratory environment where students can engage in scientific practices to investigate key concepts~\cite{podolefsky2010factors, paul2013guiding}. The design supports learners to naturally and productively ask questions, conduct experiments, discover causal relationships, reflect on results, and test their ideas, and is grounded in education research to address known student difficulties. With a collection of 174 simulations in total, 112 of which are in physics, PhET simulations are a mainstay in many physics classrooms, with over 200 million uses per year.

While it is common for instructors to use PhET simulations as part of their physics laboratory courses, few research studies have examined the impact of PhET simulations on student learning in the lab context. An early study by Finkelstein \textit{et al}. established that the interactive simulations have the potential to support some learning goals in physics labs~\cite{finkelstein2005learning}. Finkelstein \textit{et al}. found that students using the \textit{Circuit Construction Kit} simulation before the lab activity performed better on building of a real circuit, building the circuit just as fast and explaining its behavior better than the students who had used real components during the lab~\cite{finkelstein2005learning}. A handful of other studies have examined the use of PhET simulations in a lab environment, primarily studying the impact on students' conceptual understanding of the context and their inquiry skills~(e.g., Refs. \cite{abou2017effect, rustana2020analysis}).

Traditional PhET simulations are idealized using simplified physics models without inclusion of effects such as mechanical vibration, electrical noise, and measurement error. As a result, they do not model noisy, real-world situations, eliminating the need for students to consider sources of noise or error in their conceptual understanding of physics. The work described here explores how noise-enhanced PhET simulations can be used as a new tool to support student learning around measurement uncertainty in introductory physics labs.

Most physics instructors and researchers alike recognize the importance of measurement uncertainty, and learning concepts of measurement uncertainty is often one of the main learning goals of many introductory physics lab courses~\cite{kozminski2014american, pollard2021introductory}. However, there is strong evidence provided by physics education researchers that this goal is often not met, and that students often lack deep conceptual understanding of measurement uncertainty~\cite{volkwyn2008impact, day2011development, smith2019context, geschwind2024using}. To address the challenge of improving students' proficiency with concepts and practices of measurement uncertainty, we designed and developed a new PhET simulation, \textit{Projectile Data Lab} (PDL), featuring statistical noise and measurement tools in the context of projectile motion~\cite{pdlwebsite}. In addition, to allow students to easily and intuitively collect and analyze data, we integrated the \textit{Projectile Data Lab} simulation into the Common Online Data Analysis Platform (CODAP) ~\cite{codapwebsite}, creating the PDL+CODAP instructional platform. Finally, we developed lab activities that leverage the unique pedagogical capabilities of \textit{Projectile Data Lab} and the integrated PDL+CODAP instructional platform and provided instructors with guidance on how to incorporate these tools into their lab courses.

The combination of \textit{Projectile Data Lab}, the PDL+CODAP instructional platform, the associated lab activities, and the instructor facilitation work together to create a new learning environment with pedagogically-powerful learning opportunities specifically designed to enhance student proficiency with concepts and practices of measurement uncertainty. This work aims to provide a model of how noise-enhanced simulations can be designed and used in an introductory physics lab setting. Future physics education research studies are forthcoming to study how working with this new learning environment can impact students' conceptual understanding of measurement uncertainty, and how students can transfer this understanding to real-world lab settings.

It is important to preface here that we are not advocating for eliminating hands-on activities from lab courses. Labs are complex learning spaces with a myriad of goals that can be difficult to attend to all at once. To better address specific learning goals related to measurement uncertainty, we propose using these new PhET simulations to help students develop baseline proficiency in relevant skills before entering the lab. This preparation allows students to engage more productively in hands-on activities, applying and deepening their understanding of measurement uncertainty rather than struggling to learn these concepts from scratch during limited lab time.

\section{Identifying the learning goals}
To guide and structure the development of this new learning environment, we first need to identify the target learning goals for the PhET simulation, the PhET+CODAP platform, and associated lab activities around measurement uncertainty. This work occurred in the context of a broader multi-disciplinary project to identify shared learning goals between science, statistics, and data science education around data fluency, and included extensive cross-disciplinary discussions among a team of discipline-based education research faculty, experienced college and high school educators, and a data science education specialist. Here, we focus on the learning goals specifically related to the learning of measurement uncertainty in the physics context. 

The existing research related to students' conceptual challenges with measurement uncertainty, specifically research conducted in the creation of the Survey of Physics Reasoning on Uncertainty Concepts in Experiments (SPRUCE) assessment,  informed the prioritization of learning goals. As part of the development of the SPRUCE assessment, which aims to measure student learning related to measurement uncertainty in university-level physics lab courses~\cite{vignal2023survey}, the survey developers conducted interviews with 22 physics lab instructors to identify common concepts and practices related to measurement uncertainty, and their level of importance in introductory physics labs~\cite{pollard2021introductory}. They used these instructor interview results to create an initial targeted set of assessment objectives for SPRUCE that were further refined after student interviews and initial piloting of SPRUCE. These assessment objectives for SPRUCE~\cite{geschwind2024using} are included in the targeted learning goals. For the current pedagogical design work, we have not prioritized addressing ``identify actions that might improve accuracy,'' which deals with understanding of systematic rather than random error.

Beyond the SPRUCE assessment objectives, six additional learning goals are targeted. These topics were articulated by most instructors in Pollard \textit{et al}, but were excluded from SPRUCE objectives due to the difficulty of designing assessment items. These goals include ``identifying sources of error,'' as well as a number of goals related to plotting and curve fitting. With the pedagogical design features and interactive capabilities afforded by PhET simulations and a PhET+CODAP platform, we are able to include these among the targeted learning goals.

\begin{figure*}
    \centering
    \includegraphics[width=1\linewidth]{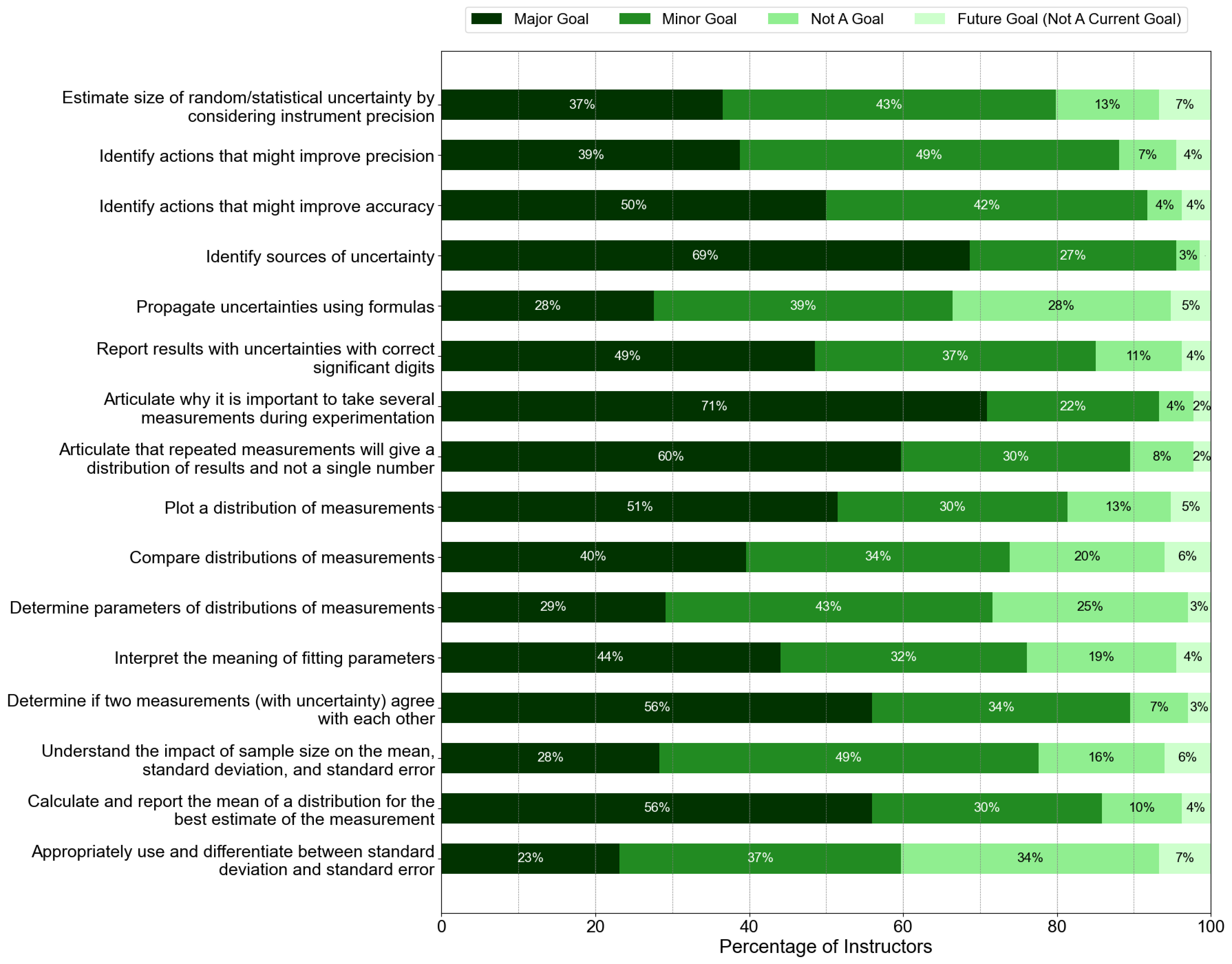}
    \caption{Percent of physics lab courses (N = 136) for which the surveyed instructors identified the importance of each targeted learning goal as a major goal (dark green), minor goal (green), not a goal (lime green), and future goal (light green). The items are listed in the same order as they appear in Fig. \ref{fig2}.}
    \label{fig1}
\end{figure*}

To further confirm that these targeted learning goals align with physics instructors' course goals related to measurement uncertainty, we surveyed college-level physics lab instructors and coordinators who had previously expressed interest in using noise-enhanced PhET simulations in their labs, and received responses for 136 physics courses. Almost all of the instructors teach introductory lab courses, and they would like to apply these PhET simulations in their courses in a variety of ways: as pre-lab activities, classroom demonstrations, a supplement to current hands-on labs, or a replacement for current hands-on labs. When asked about the extent to which the targeted learning goals align with their course objectives, most surveyed instructors reported that all of the outlined goals were consistent with their own. The distribution of their responses are plotted in Figure \ref{fig1}, with all goals being a major or minor goal for the majority of instructors and 7 of 16 goals being a major goal for the majority.

\begin{figure*}
    \centering
    
    \label{fig2}
    \includegraphics[width=1\linewidth]{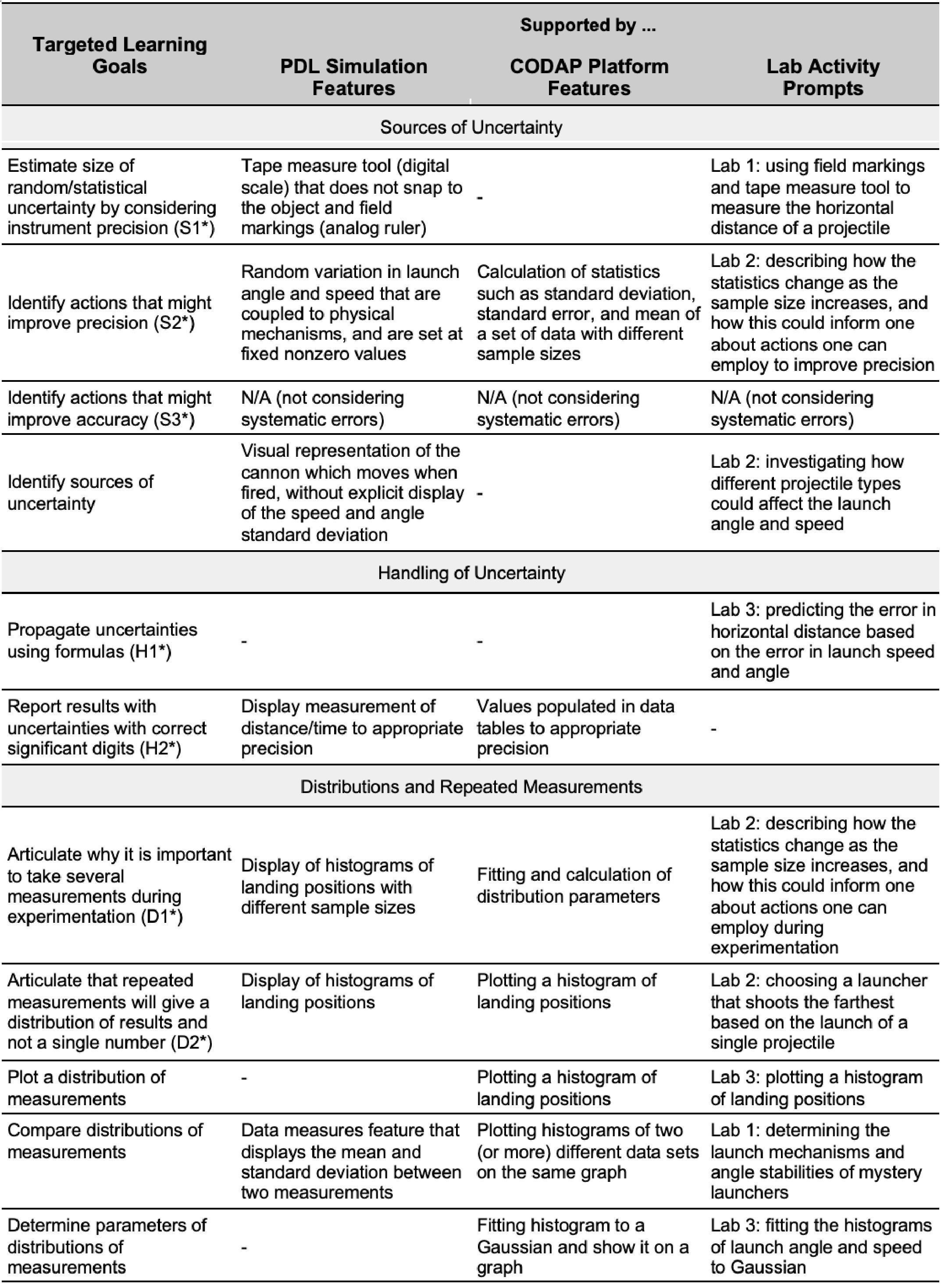}
    
\end{figure*}

Support for these targeted learning goals is provided by the pedagogical design and scaffolding of the three instructional tools working together: the \textit{Projectile Data Lab} simulation, the PDL+CODAP instructional platform, and the structured lab activity worksheets. The simulation's design includes implicit scaffolding, and feedback that work synergistically with CODAP's data analysis and visualization capabilities, shaping the pedagogical possibilities, structure, and prompt design within the lab activities. These elements form an integrated, technology-enhanced learning environment that distributes the scaffolding responsibilities~\cite{kaldaras2024employing}: the embedded, implicit scaffolding within the technology tools and the explicit scaffolding within the lab activity collectively scaffold student learning of each goal. Figure \ref{fig2} details the design features of the PDL simulation, CODAP platform, and lab activities, and how they work together to support the targeted learning goals.

\begin{figure*}
    \centering
    \includegraphics[width=1\linewidth]{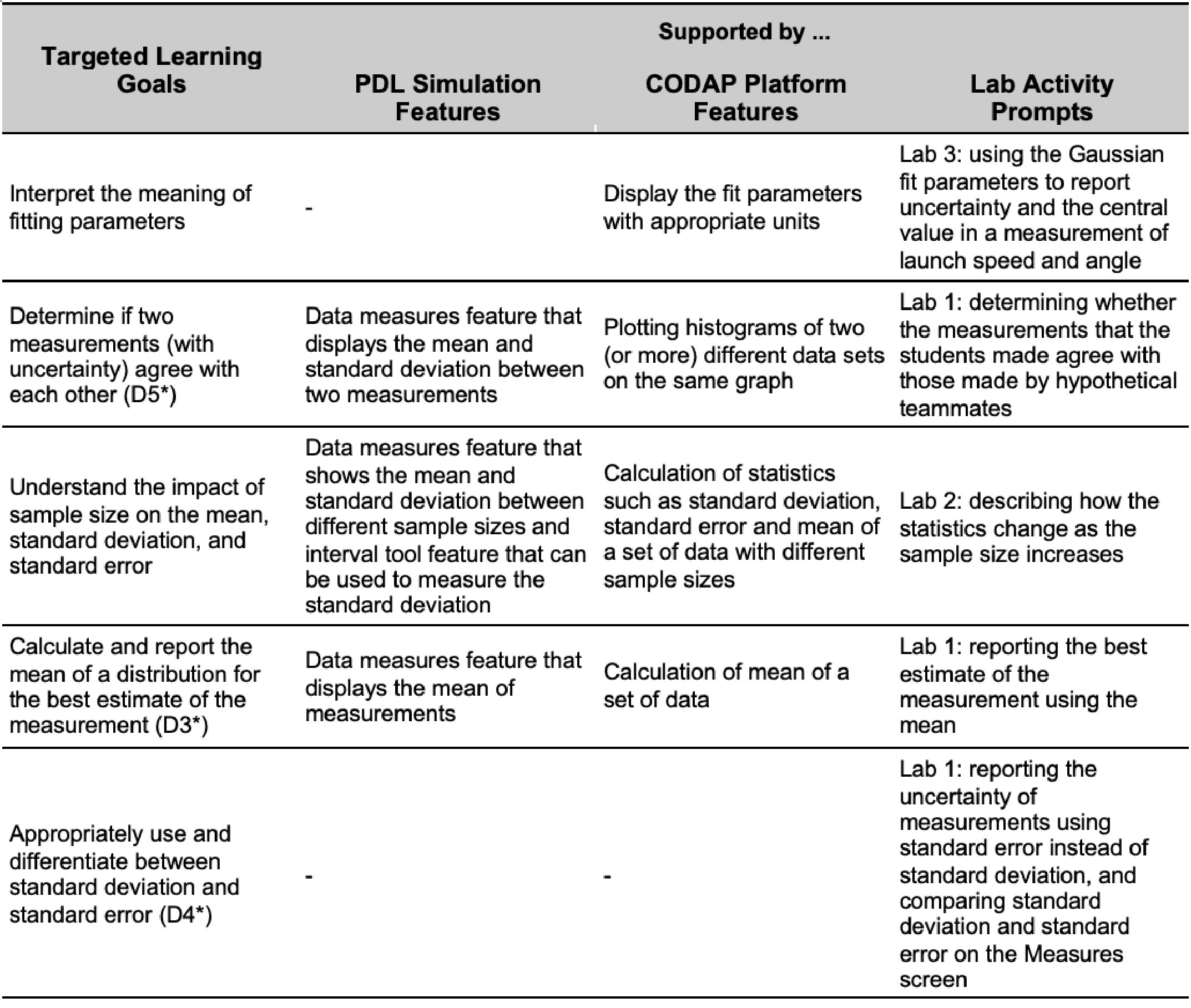}
    \caption{Targeted learning goals, along with associated capabilities and design features of the PDL simulation, CODAP platform, and lab activities that support these learning goals. If the target learning goal is one of the SPRUCE assessment objectives, the corresponding label is included in parenthesis with an asterisk marked at the end~\cite{geschwind2024using}.}
\end{figure*}

\section{Overall philosophy of designing the learning environment}
To promote learning of measurement uncertainty concepts and practices using the new learning environment, we drew inspiration from constructivist learning theory, physics education research on laboratory instruction, and PhET's research and experience on designing simulations and simulation-based learning environments.  Here, we highlight three key guiding principles that we employed in the design of the PhET simulation, the integrated PhET+CODAP platform, and the associated lab activities.

\subsection{Creating opportunities for open-ended exploration}
Several studies have shown that traditional ``cookbook'' labs that feature only guided activities often do not contribute to the improvement of student conceptual understanding beyond lectures, yet they have the potential to negatively impact students' views of experimental physics~\cite{wieman2015comparative, holmes2018introductory}. On the other hand, open-ended lab activities, where students are provided with the opportunity for more agency in making decisions about the experiments, are shown to improve students' conceptual understanding, as well as their views of experimental physics~\cite{holmes2018introductory, wilcox2016open, walsh2022skills}. As such, we aim to create simulation features and lab activities that are more open-ended in nature to provide students with the opportunities to have agency~\cite{holmes2020developing, kalender2021restructuring}. For instance, each of the three lab activities that we designed for the simulation starts with five minutes of open exploration, during which students explore any and all aspects of the simulation without explicit directions. This open exploration not only familiarizes students with controls and decreases the need for instruction later in the lab, but also supports student agency and ownership of their learning process. Students can feel empowered as they discover simulation features without being told about them directly~\cite{paul2013guiding}.

\subsection{Scaffolding the overall learning experience}
Even though we are moving away from fully-guided lab activities, giving students complete autonomy with little to no instruction can also lead to frustration and confusion~\cite{kalender2021restructuring, henige2011undergraduate}. As such, we scaffold the overall learning experience in ways that promote student agency and productive engagement around the learning goals~\cite{podolefsky2010factors, paul2013guiding}. For instance, after students engage in open play with the simulation, we scaffold the remainder of the lab activities in accordance with the process of ``predict,'' ``gather and analyze data,'' and ``draw conclusions and reflect,'' to not only facilitate the construction of new knowledge around measurement uncertainty, but also to support engagement with authentic scientific practices~\cite{aaptrecommendations}. For all three lab activities, students are provided with the opportunity to make their own predictions of the simulation outcome based on their prior knowledge, test their predictions in the simulations, and either confirm or revise their predictions and reasoning using the experimental results and multiple representations in the simulation.

\subsection{Employing the contrasting cases approach}
The contrasting cases approach uses carefully designed examples to draw students' attention to the features of a certain concept, and has shown to be effective at improving student learning~\cite{schwartz2011practicing, roelle2015effects}. For the simulation and lab activities, we provide students with the opportunity to evaluate measurement uncertainties in different scenarios and compare them side-by-side, so that their understanding of measurement uncertainty is less context dependent. For instance, when defining the launch speed associated with different launch mechanisms (spring, pressure, and explosive) of the launcher, the pressure launch mechanism is associated with a higher mean yet a lower standard deviation of launch speed as compared to the spring, in an effort to challenge students' perception of correlation between the mean and standard deviation, while purposely tying the launcher behaviors to more physically realistic, messier similarities and differences of overlapping properties. As such, changing a launcher can affect several parameters at once (mean and standard deviation of launch speed and angle), so students must design controlled experiments rather than just sliding a single control to change these parameters. This makes the contrasting cases richer and more puzzle-like. Additionally, the launchers were carefully sequenced in the simulation, while some differ only in mean speed, and others only differ in the standard deviation of speed, etc. This mix creates meaningful similarities and differences for students to investigate.

\section{Designing the PhET simulation with noise}
  
The simulation design employed PhET's established design principles and development processes for creating effective educational simulations~\cite{podolefsky2013implicit}. PhET simulations share a few key characteristics. Each simulation presents an open exploratory environment, with an intuitive interface that supports student-driven exploration and discovery through interactivity, immediate feedback, and multiple dynamic representations. The result is a highly flexible tool that can support multiple instructional contexts and accommodate different teaching approaches and learning goals. With this framing and with this work situated in a broader data fluency context, the interdisciplinary simulation design team - including pedagogical and content experts from science, statistics, and data science, software engineers, and simulation design experts - used a structured design process. 

The simulation design team first needed to decide on a contextual topic that could best support experimentation and conceptual understanding development around each of these targeted learning goals for physics. Projectile motion was quickly identified for its strong potential, as it allows for multiple trials and is a natural fit for layering uncertainty into the launch parameters for the projectile (such as launch speed, launch angle, etc.), creating a resulting distribution of landing positions. While PhET's collection includes the popular \textit{Projectile Motion} simulation, the team found that merely adding uncertainty options to the existing simulation failed to address some of the most central learning goals, particularly around identifying sources of uncertainty, and hampered our ability to implicitly scaffold students' interaction towards the development of these specific learning goals.

Thus, the team designed a brand new PhET simulation, \textit{Projectile Data Lab},  which centers on student exploration of systems with uncertainty, the representation of variability, and the connection of physical apparatus with both the source and magnitude of variability.  Implicit scaffolding is employed throughout the design to support productive student exploration~\cite{podolefsky2013implicit}. 

\begin{figure}[t]
  \includegraphics[width=1\linewidth]{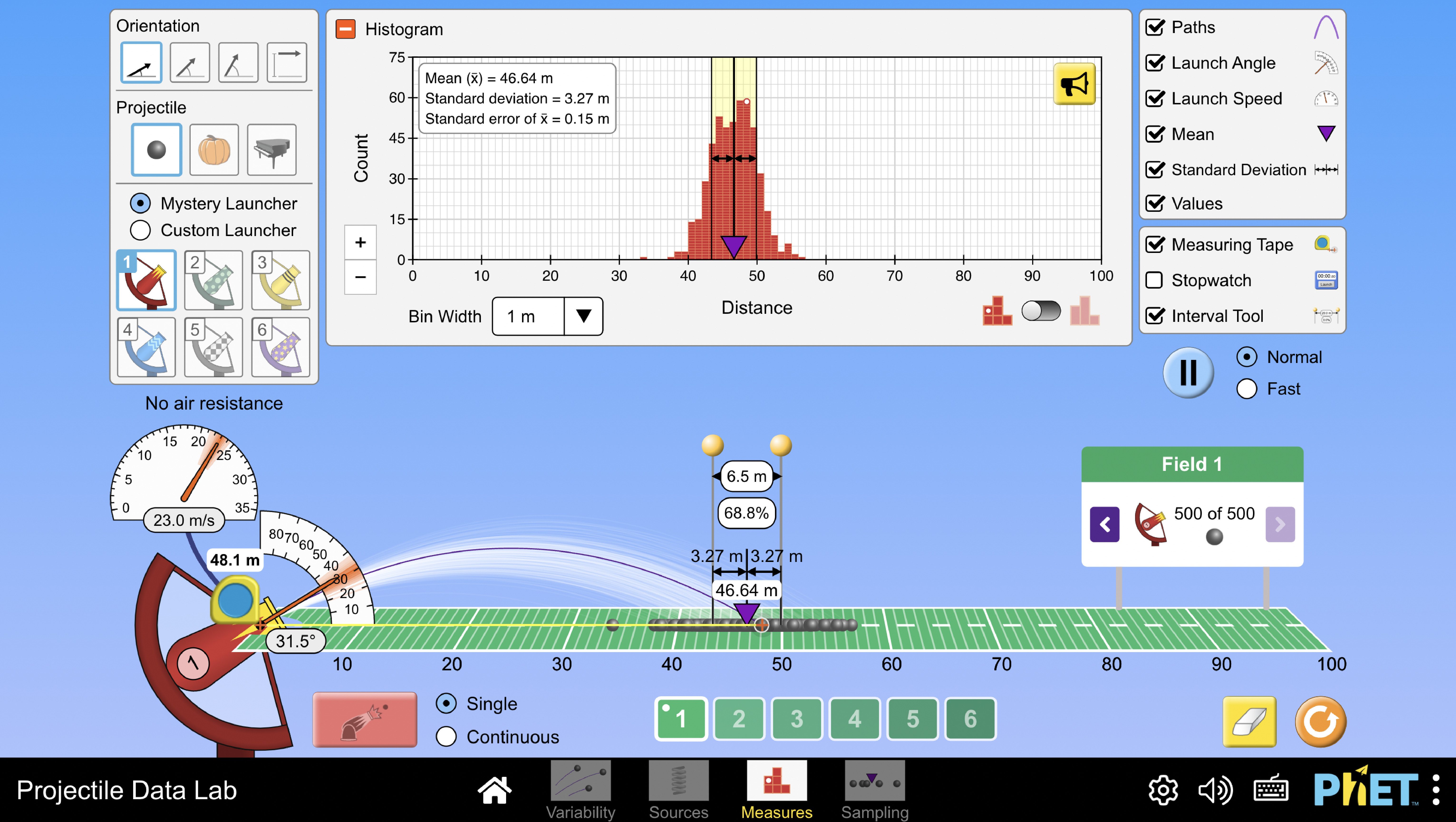}
  \caption{The `Measures' Screen of the \textit{Projectile Data Lab} PhET simulation. Students can choose from six different launchers with varying amounts of statistical noise in their launch speed and angle. The variation in the projectiles' horizontal distances can be visualized in the histogram panel. The interval tool allows the students to  quantify the amount of data in a particular range.\label{fig3}}
\end{figure}

The \textit{Projectile Data Lab} features four screens -- Variability, Sources, Measures, and Sampling -- that serve to scaffold student exploration and build upon experience and understanding in earlier screens. The first screen includes the fewest features, with each successive screen layering on new capabilities or representations. This approach reduces students' initial cognitive load and supports students productive exploration and construction of knowledge structures on the topic of measurement uncertainty. Across all screens, students can control the pace of their data collection, with the ``fast,'' ``continuous'' launching of projectiles enabling rapid exploration and inquiry cycles. An example interface of the simulation is shown in Figure \ref{fig3}. For detailed annotations of the simulation features on each screen, readers can refer to the Teacher Tips document for \textit{Projectile Data Lab}~\cite{pdlwebsite}.

In the first screen, `Variability', students can explore and compare the distributions of projectiles fired with six different launchers, choosing among three different projectile types and four different launcher orientations, creating a set of contrasting cases. The mystery launchers are specifically designed with different amounts of random noise in their launch speed and angle, leading to variation in the projectiles’ paths and horizontal distances when they land on the field. Students can also switch between six independent fields for their data collection, providing them with control and flexibility over how they collect and compare data with different conditions. This design presents students with data that have significant variability, implicitly encouraging them to explore this variability and supporting them to collect and organize their data using different fields as they investigate the difference in variability for each launcher. To further support students' exploration, the `Variability' screen also features tools to measure the variation in launch speed and angle of projectiles, which include a novel heat map representation of the variability in launch speeds and angles, as well as a measuring tape and stopwatch to mimic real-world measurements of distance and time of flight with their inherent uncertainties. A histogram representation plots the distribution of the horizontal distances of the projectiles on the current field, and allows students to adjust bin width and toggle between displaying stacked individual data blocks or a single bar per bin. Finally, a data selection tool allows students to easily review each data point, coordinating the launcher, the orientation, the projectile type, the landed projectile, and histogram point. Each of these simulation features encourages students to explore various aspects of this physical setting without explicit guidance.

The `Sources' screen allows students to delve deeper into the physical mechanisms behind the variability in the launch properties of the projectiles. Students can create their own custom launchers, selecting one of the three launch mechanisms (spring, pressure, and explosive) and setting the angle stability (min to max), and see ``inside'' as the inner-workings of the launcher change. The launch mechanisms are categorical properties of the launcher that affect the average and standard deviation of launch speed, whereas the angle stabilizer represents the physical mechanism that affects standard deviation of launch angle. The purpose of this screen is to present students with the ability to alter physical properties of the launcher in a more transparent fashion, encouraging the students to think about the sources of uncertainty. Rather than treating the launcher as a ``black box,'' students can now design controlled experiments to investigate how the launch mechanisms and angle stabilizer determine the variability of the resulting projectile motion.

The `Measures' screen creates a sandbox environment that allows students to compare the mystery launchers from the `Variability' screen and the custom launchers from the `Sources' screen with additional tools and visualizations to aid their analysis on quantifying the variability. This screen includes several new features including several statistical data measures, as well as detailed quantitative data. Students can  graphically display the mean and standard deviation of the projectiles on the field and on the histogram, and watch these dynamically update as more projectiles are launched. Displaying the values supports students to begin connecting their conceptual and quantitative understanding of these measures, with quantitative values for mean, standard deviation, and, if selected in the Preferences menu (gear icon), the standard error. A new Interval Tool allows students to quantify the percentage of projectiles that landed in a particular range on the field. This screen allows students to quantify the variability in launch speed and angle of both the mystery launchers and custom launchers, and to explore the inner workings of the mystery launchers. In doing so, students are prompted to reflect on how one quantifies and compares data with variability using different statistical measures.

The `Sampling' screen introduces students to more advanced statistical concepts and techniques, including data sampling, sampling distributions, and standard error. Students can launch cannonballs in clusters by selecting a sample size (n = 2, 5, 15, or 40). Each sample is fired in rapid succession from a mystery launcher, and once all cannonballs have landed, the mean horizontal distance for that sample is marked on the field. Unlike the first three screens, the `Sampling' screen focuses more explicitly on statistics rather than scientific practices. Because it covers content not typically emphasized in science courses, it is not utilized in the physics lab activities.


During the course of its development, the \textit{Projectile Data Lab} simulation was tested with student interviews, where students openly engage with the PhET simulation and discuss their reasoning aloud. These interviews aim to test whether the simulation interactions are intuitive for students with little to no explicit directions, examine how students are or are not productively engaging with the learning environment, and assess whether this new simulation is properly scaffolded. Insights obtained from these interviews informed the refinement of the simulation, and this process was iterated until interviews showed that the simulation worked as intended. The list of simulation features corresponding to each targeted learning goal is presented in Figure \ref{fig2}.

\section{Integrating the simulation with CODAP}

Driven by the learning goals associated with data analysis, plotting, and curve fitting, we integrated the \textit{Projectile Data Lab} simulation with CODAP~\cite{codapwebsite} to develop the PDL+CODAP instructional platform that allows students to easily analyze data collected in the simulation. Developed by Concord Consortium, CODAP is a free, browser-based, open-source software that allows for collection, organization, and display of multiple data representations, and dynamically links these representations such that data selected in a table is also highlighted across the other graphs and tables where it appears. This ``linked selection'' capability can assist students in making sense of the data. While PhET simulations can provide a context for students to conduct experiments and generate data, CODAP has proven to be a robust analysis tool for connecting to various experimental data sources for lab use~\cite{gresalfi2020interdisciplinarity}.

To be able to implement the types of analysis useful for exploring measurement uncertainty with distributions typically used in the physics lab, we added new functionalities into CODAP. Students can now display the standard error for a distribution,  and adjust the number of standard errors represented by the bar (a feature more often used by the statistics community). As an advanced feature, accessible only via a query parameter, students can plot the results of a least-squares fit of a Gaussian function with the histogram along with the calculated fit parameters and uncertainties. 

The integration of the \textit{Projectile Data Lab} simulation  with CODAP relies on PhET's advanced PhET-iO technologies. PhET-iO simulations are fully-instrumented versions of PhET simulations that support customization, API integration, and back-end data collection~\cite{moore2018advances, phetiowebsite}. The design of the embedded simulation-CODAP interface has been carefully crafted, and wraps the simulation interface with a side-panel where students need to explicitly choose which data they want to collect on each launch from a list of setup and measurement variables. For each projectile launched, the API integration is used to transfer the selected data to the Simulation Data table.  As in the real world, if the data are not collected at the time of launch, they are not available for analysis.

\begin{figure}[t]
  \includegraphics[width=1\linewidth]{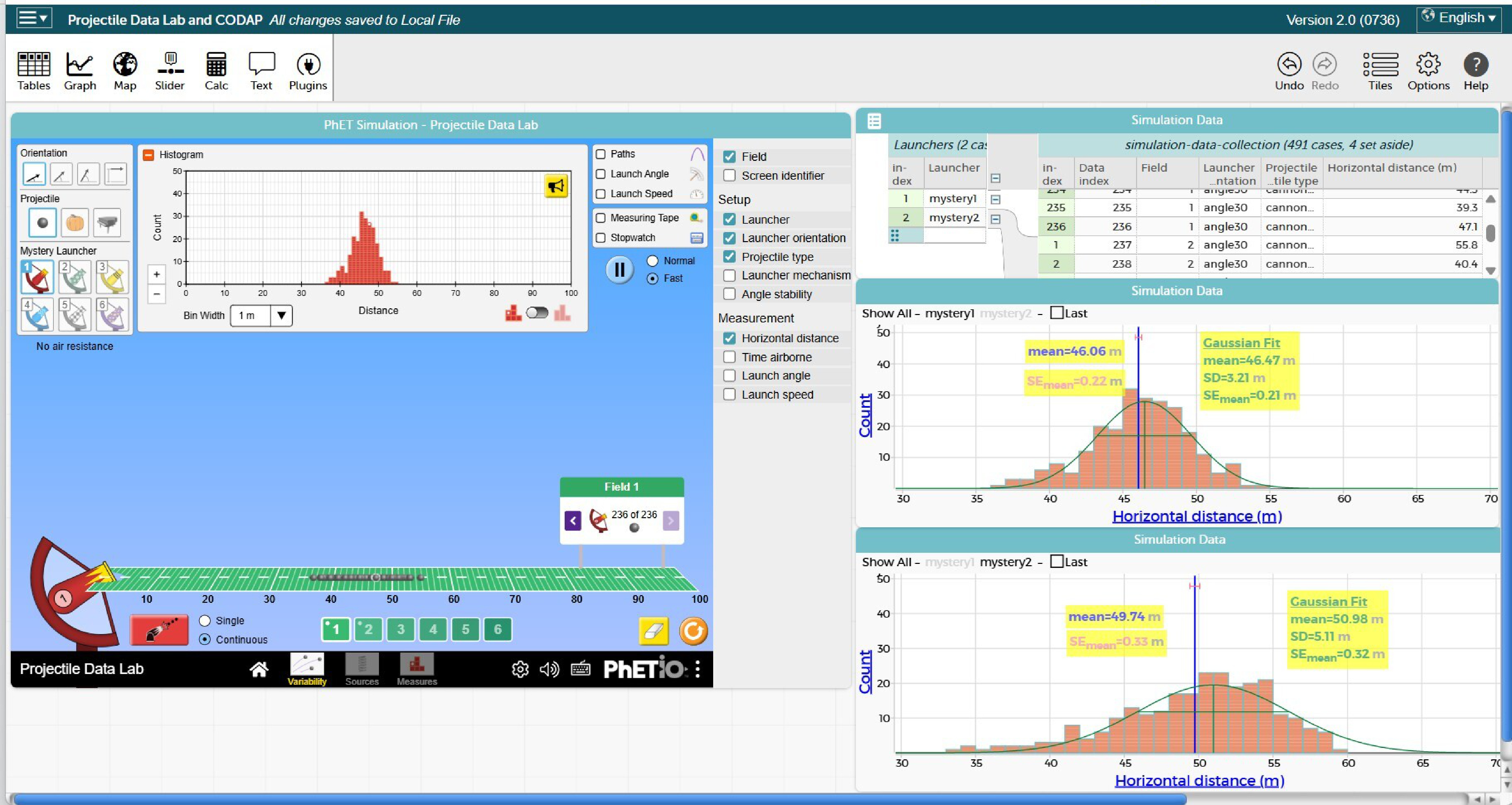}
  \caption{An example interface of the integrated PDL+CODAP instructional platform. Students can choose the setup and select the data they wish to collect. As they launch projectiles in the simulation, the selected data automatically populates in a table. Using a simple drag-and-drop interface, students can plot variables and explore relationships. CODAP also enables comparisons between distributions and provides tools for calculating statistics across selected variables.\label{fig4}}
\end{figure}

The resulting integrated PDL+CODAP instructional platform allows students to take data in the simulation, then visualize and analyze these data using CODAP. An example interface of this CODAP-integrated instructional platform is shown in Figure \ref{fig4}. Students need to first decide how to organize their experiments, select what data to collect, and begin analysis. They can create plots and analyze data, by dragging variables from the Simulation Data table to the Simulation Data graph.

Similar to the \textit{Projectile Data Lab} simulation, the integrated PDL+CODAP instructional platform has been tested with student interviews, which revealed some challenges with the user interface and pedagogical scaffolding of features for the learning goals. Revisions were made based on these student interviews, and further interviews established that the desired engagement is being achieved. The full list of CODAP features corresponding to each targeted learning goal is presented in Figure \ref{fig2}.

\section{Designing the associated lab activities}
Leveraging the affordances provided by the \textit{Projectile Data Lab} simulation and the PDL+CODAP instructional platform, we designed three paired lab activities with the aim of helping students develop and practice their skills around measurement uncertainty. These lab activities are not intended to replace hands-on activities from labs, but rather to supplement them by enhancing students' conceptual understanding of measurement uncertainty and familiarizing them with common experimental practices.

Each paired lab activity focuses on developing student conceptual understanding of some components of measurement uncertainty, guided by a thematic subset of the targeted learning goals with each lab building on the previous one. Lab 1  focuses primarily on goals related to single measurements. Lab 2 delves more deeply into the topic of multiple measurements. In Lab 3, students  ``determine parameters of distributions of measurements,'' focusing on understanding how multiple measurements improve the reliability and goodness of fit. 

Instructors can use these three lab activities in sequence, with each one becoming progressively more advanced and scaffolded according to the complexity of the learning goals. However, each lab is also designed to be self-contained and modular, allowing instructors to use one or two activities independently. This flexibility is especially useful when certain learning goals do not align with a particular course, and the instructor wants students to focus on a specific subset of skills or concepts. Finally, to encourage students to consider measurement uncertainty concepts in the context of objects in the real world, each of the lab activities are grounded in a scenario that is likely relevant to the student lives. The contexts are depicted in the form of short stories where the main character is the student, and the problem statements in the activities include plausible motivations or reasons for the student to perform experiments. This approach further emphasizes the need for students to make their own decisions by employing their knowledge of measurement uncertainty.

Lab 1 focuses on determining the measurement uncertainty of a single measurement and of multiple measurements~\cite{pdllab1}. It is situated in a context of a company that is developing and testing out their latest projectile launchers, the student is hired as a member of the research and development team. The first part of this activity involves students measuring the horizontal distance of a single projectile for different projectile types from the mystery launchers using the field markings and measuring tape tool in the \textit{Projectile Data Lab} simulation, and subsequently reflecting on how the projectile size and instrument precision influenced their measurements and associated uncertainties. The second part of this activity involves the student quantitatively investigating the launch angle, launch speed, and horizontal distance of the projectile for a mystery launcher of their choice using the PDL+CODAP-platform, building their own custom launcher that mimics the behavior of the mystery one, and finally reflecting on how statistics such as standard deviation and standard error are used to make comparisons and conclusions.

Lab 2 focuses on comparing measurement uncertainties between single and multiple measurements~\cite{pdllab2}. It is situated in a context where the student is entering a university-level science competition, and the task is to launch the projectile as far as possible. Students are first tasked to take multiple measurements using a mystery launcher of their choice, and use CODAP to plot the relationship between different statistics (mean, standard deviation, and standard error) and sample size. Based on this information, they are asked to form a strategy on determining the mystery launcher and projectile type such that the projectile travels as far as possible, but they are allowed to launch only one projectile for the competition. The activity ends with a reflection on the probabilistic nature of random error and the importance of taking multiple measurements during experimentation.

Lab 3 focuses on more advanced topics, such as error propagation and curve fitting~\cite{pdllab3}. It is situated in a context where the student is a special-effects coordinator for an upcoming movie, and the goal is to set up a projectile launcher and camera to get a close-up recording of a cannonball landing. The first part of this activity involves the student predicting the uncertainty of horizontal distance of a projectile based on the uncertainty of launch speed and time airborne. However, since the range equation for projectile motion involves trigonometric functions, this may be particularly challenging for students with limited mathematical background on partial derivatives. To remove this barrier, the activity fixes the launch angle at zero, so that with the small-angle approximation, the range equation transforms to a linear form. The second part of this activity involves fitting the horizontal distance distribution to a Gaussian function using CODAP, and estimating the probability of the projectile landing within a certain range. 

Each of these lab activities has been tested with student interviews, where students worked through each of the activities designed for the \textit{Projectile Data Lab} simulation and the PDL+CODAP instructional platform, discussing their reasoning aloud. Based on the interview observations, we refined and improved the activity design, such that they are ready to be adapted by any interested instructors in their classrooms. The full list of lab activities corresponding to each of the targeted learning goals is presented in Figure \ref{fig2}.

\section{Conclusions and future work}
Motivated by the common focus of many introductory lab courses on student learning of measurement uncertainty concepts and practices, we designed, developed, and tested the \textit{Projectile Data Lab} simulation, the first PhET simulation to incorporate statistical noise, answering a long-standing request of physics instructors. Moreover, we pioneered the integration of PhET simulations with CODAP, creating the PDL+CODAP platform and providing a new easy-to-use data collection and analysis solution. Finally, we developed and tested three related lab activities, all of which are focused on developing understanding of measurement uncertainty concepts and practices. \textit{Projectile Data Lab}, the integrated PDL+CODAP platform, and the collection of their associated lab activities are all freely available.  Since publication in May 2024, the\textit{Projectile Data Lab} simulation has been run over 850,000 times, and translated into 35 languages.

This work demonstrates how research-based design of noise-enhanced PhET simulations and an integrated PhET+CODAP environment can open up powerful new pedagogical opportunities with the potential to transform how we develop students' conceptual understanding around measurement uncertainty and prepare them to engage in real-world experiments in introductory physics lab courses. Noise-enhanced simulations can provide a natural setting for students to practice their experimental skills on handling variability, by adding authentic and contextualized sources of variability to experimental equipment and by including measurement tools that mimic the measurement uncertainty of real-world lab instruments. These simulations allow students to engage in rapid inquiry cycles and to dynamically explore different size data sets. A noise-enhance simulation integrated into the CODAP platform can support a breadth of exploration that may not be possible or practical with real lab equipment.

The design strategies and technological approaches utilized here for \textit{Projectile Data Lab} can serve as a model for adding noise and sources of uncertainty to more simulations in physics and across the sciences. Similarly, the strategies used to create the integrated PDL+CODAP platform, could be leveraged by other education technology platforms and physics curriculum  (e.g., Pivot Interactives~\cite{pivotwebsite}, TUVA~\cite{tuvawebsite}, BrainPop~\cite{brainpopwebsite}) to teach key concepts in measurement and measurement uncertainty. With these new tools being available online, a suite of noise-enhanced simulations and CODAP-integrated environments could bring new learning opportunities to millions of students around the world, including in many institutions that lack access to physical laboratory equipment.

Noise-enhanced simulations have limitations and are not well-matched for all learning goals. In physics lab courses, there are important cognitive tasks common in hands-on lab activities that are not addressable with these simulations, such as troubleshooting equipment and awareness of lab safety. In addition, the addition of statistical noise adds additional complexity and cognitive load for students, and as such, is not the best instructional tool for students who are trying to understand the underlying physics concepts. Therefore, we recommend instructors use noise-enhanced simulations like \textit{Projectile Data Lab} when their objectives are aligned with the targeted learning goals associated with measurement uncertainty. And again, we recommend their use only as a supplement to hands-on lab experiences.

Finally, future research will utilize additional think-aloud student interviews and validated classroom assessment to study the impact of the \textit{Projectile Data Lab} simulation, the PDL+CODAP platform, and the lab activities on student understanding of measurement uncertainty concepts and practices. In addition, future development work will focus on applying the design patterns and technologies here to new contexts, with the ultimate goal of creating a suite of ``Data Lab'' variations of PhET simulations.

\end{document}